\documentclass{article}

\title{
\vspace{1cm} $\rho-\omega-$Interference in $J/\psi-$Decays and
$\rho\rightarrow \pi^+\pi^-\pi^0$ Decay \footnote{Supported by
National Natural Science Foundation of China (90103002)} }
\author{Liu Fang$^{1,2}$ \footnote{E-mail address: liuf@mail.ihep.ac.cn}, Li Jin$^{2}$,Yi-Bin Huang$^{1}$,
Mu-Lin Yan$^{1}$\footnote{Corresponding author; E-mail address:
mlyan@staff.ustc.edu.cn}\\
1 Interdisciplinary Center for Theoretical Study,
University of Science and\\
Technology of China Hefei, Anhui 230026, P.R. China\\
2 Institute of High Energy Physics, Beijing 100039,
 P.R.China }
\date{}

\begin{document}
\maketitle

\begin{description}
\normalsize \item [~~~~~~Abstract]We study
$\rho-\omega-$interference by analyzing $J/\psi\rightarrow
\pi^+\pi^-\pi^0\pi^0$. PDG-2002 data on $J/\psi$ decays into $PP$
and $PV$ ($P$ denotes pseudoscalar mesons; $V$, vector mesons) are
used to fit a generic model which describes the $J/\psi$ decays.
From the fits, we obtain anomalously large branching ratio
$Br(\rho^0\rightarrow \pi^+\pi^-\pi^0)\sim 10^{-3}-10^{-2}$. A
theoretical analysis for it is also provided, and the prediction
is in good agreement with
 the anomalously large $Br(\rho^0\rightarrow \pi^+\pi^-\pi^0)$. By
 the fit, we also get the $\eta-\eta'-$mixing angle
 $\theta=-19.68^o\pm 1.49^o$ and the constituent quark mass ratio
 $m_u/m_s\sim 0.6$ which are all reasonable.

\vskip 4mm \item[~~~~~~Keyword] $J/\psi-$decays, $\rho - \omega$
interference, SU(3)-breaking effect, $\eta - \eta^{\prime}$ mixing
\end{description}

\section{introduction}
\vskip 3mm \normalsize

~~~~Recently it has been predicted in ref.$^{\cite{WJY}}$ that
there is a large isospin symmetry breaking enhancement effect in
the decay $\rho^{0} \rightarrow \pi^{0} \gamma$ comparing with
$\rho^{\pm} \rightarrow \pi^{\pm} \gamma$ due to
$\rho-\omega$-interference, which was called as hidden
isospin-breaking effects in $^{\cite{WJY}}$. This prediction has
been confirmed by the renewed data in PDG-2002$^{\cite{PDG}}$.
Following the discussion of $^{\cite{WJY}}$, it could be expected
that a similar large hidden isospin-breaking effect should also
exist in $\rho^{0} \rightarrow 3 \pi$ due to $\rho - \omega$
interference. In this paper we try to analyze $J/\psi\rightarrow
(\rho^0, \omega) (\pi^o,\eta,\eta') \rightarrow (3\pi)
(\pi^0,\eta, \eta')$, and to reveal $\rho-\omega$-interference
effect and then finally to abstract out the branching ratio of
$\rho^{0} \rightarrow 3 \pi$. Such an analysis should be necessary
for further confirming the enhancement effects mentioned above,
and be also interesting for the $G$-parity violating process
studies.

It is very difficult to directly measure $B_r(\rho^0\rightarrow
 3\pi)$ experimentally both because $\Gamma_\rho >> (m_\omega-m_\rho)$ and because
the $G$-parity conserving decay mode of $\omega\rightarrow 3\pi$
is dominate. This is the reason why there is still no a reliable
value for $B_r(\rho\rightarrow 3\pi)$ yet so far in the
literature$^{\cite{PDG}\cite{Vas}}$. Fortunately, this quantity
can be obtained by fitting the data of $J/\psi\rightarrow PP$ and
$PV$ (where $P$ denotes the pseudoscalar meson nonet $(\pi, K,
\eta, \eta')$, and $V$, the vector mesons $(\rho, \omega, K^*,
\phi)$)$^{\cite{Bramon}}$. Actually, more than fifteen years ago,
$B_r(\rho^0\rightarrow 3\pi)$ was estimated in
ref.$^{\cite{Bramon}}$ by using the MARK-III data of $(J/\psi
\rightarrow PP,\;PV) $-decays with $P={(\pi, K)}$ and $V={(\rho,
\omega, K^*)}$. However, people including the authors of
ref.$^{\cite{Bramon}}$ does not think  their result of
$B_r(\rho^0\rightarrow 3\pi)$ is very reliable  (see the
discussion in ref.$^{\cite{Bramon}}$ and ref.$^{\cite{PDG}}$). The
reasons are multi-ply, and some of them may be as follows: 1, the
$J/\psi$-decay data quality at that time was not good enough ; 2,
the fitting is not complete because the processes of ($J/\psi
\rightarrow \eta V,\; \eta' V,\; P\phi$) were not considered; 3,
lacking a theoretical understanding why their result is so
significantly different from the result of ref.$^{\cite{Vas}}$
which is quoted by PDG$^{\cite{PDG}}$. Our motive of this paper is
to try to solve those problems: 1, We shall use nowadays data of
$(J/\psi \rightarrow PP,\;PV)$ in PDG-2002$^{\cite{PDG}}$ to
perform the fit; 2, ($J/\psi \rightarrow \eta V,\; \eta' V,\;
P\phi$) will be considered in our analysis; 3, And the rationality
of the result will be argued, i.e., we'll see that the result is
just consistent with the theoretical analysis in
ref.$^{\cite{WJY}}$.

The contents of the paper are organized as follows. In section 2
we describe the $\rho-\omega-$interference in the process $J/\psi
\rightarrow \pi^+\pi^-\pi^0\pi^0$, and give the the branching
ratio formulas for ($J\psi \rightarrow PP,\;PV$). In section 3, by
using the PDG-2002 $J/\psi$ decay data and the formulas given in
the section 2 we perform the datum fits. The $Br(\rho^0\rightarrow
\pi^+\pi^-\pi^0)$ is obtained. The section 4 is devoted to
estimate $Br(\rho^0\rightarrow \pi^+\pi^-\pi^0)$ by theoretical
analysis. Finally, we briefly discuss the results.

\section{$\rho - \omega$ Interference and branching ratios of $J/\psi$ decays into PV and
PP}
\vskip 3mm
\normalsize
~~~~The elusive $\rho^0 \rightarrow
\pi^+\pi^-\pi^0$ decay could be observed in $J/\psi$ decays into
the $\pi^+\pi^-\pi^0\pi^0$ final state$^{\cite{Bramon}}$. Indeed,
this decay can proceed through the interfering channels
$J/\psi\rightarrow\omega\pi^0 \rightarrow \pi^+\pi^-\pi^0\pi^0$
and $J/\psi\rightarrow\rho \pi^0 \rightarrow
\pi^+\pi^-\pi^0\pi^0$. Because $J/\psi\rightarrow\rho \pi^0$ is
caused both by strong interaction via 3 gluons and by
electromagnetic (EM) interaction and $J/\psi\rightarrow\omega
\pi^0$ is caused by EM interaction merely, the passibility of the
decay $J/\psi\rightarrow\rho \pi^0$ is much larger than one of
$J/\psi\rightarrow\omega \pi^0$, i.e., $\Gamma
(J/\psi\rightarrow\rho \pi^0)>>\Gamma (J/\psi\rightarrow\omega
\pi^0)$ (by using the 2002-PDG data$^{\cite{PDG}}$ we have $\Gamma
(J/\psi\rightarrow\rho \pi^0)\simeq 2.5\times 10^{2}\Gamma
(J/\psi\rightarrow\omega \pi^0)$). Consequently, even though
$\Gamma (\omega \rightarrow\pi^+\pi^-\pi^0)$ may be much larger
than $\Gamma (\rho \rightarrow\pi^+\pi^-\pi^0)$, it is still
hopeful to measure $\Gamma (\rho \rightarrow\pi^+\pi^-\pi^0)$ by
studying the $\rho-\omega$-interference effects in the process of
$J/\psi \rightarrow \pi^+\pi^-\pi^0\pi^0$.

To $J/\psi \rightarrow (\rho,\omega) \pi^{0} \rightarrow 3 \pi
\pi^{0}$, the corresponding s-dependence Breit-Wigner is written
as$^{\cite{Bramon}}$ $F(s)\equiv BW_{\omega}(s)+\epsilon e^{i
\theta'}BW_{\rho}(s)$,where
\begin{equation}
\label{3}
BW_{i}=\sqrt{\frac{m_{i} \Gamma_{i}}{\pi}} \frac {1}{m^{2}_{i}-s-i m_{i}
\Gamma_{i}}
\end{equation}
is the normalized Breit-Wigner curve for $i=\omega$, $\rho$
resonance with the mass of $m_{i}$ and total width of
$\Gamma_{i}$. The factor $\epsilon(\theta')$ is the modulus(phase)
of the amplitude proceeding through the $\rho$ resonance relative
to $\omega$ resonance,which can be written as:
\begin{equation}
\label{e4} \epsilon=\frac{|A(J/\psi \rightarrow \rho^{0}
\pi^{0})|}{|A(J/\psi \rightarrow \omega \pi^{0})|} \times
\sqrt{\frac{BR(\rho^{0} \rightarrow  3 \pi)} {BR(\omega
\rightarrow 3 \pi)}}
\end{equation}
and $\theta'=\theta_{(J/\psi \rightarrow i \pi^{0})}+\theta_{(i
\rightarrow 3\pi)}$ with $i=\rho,\omega$.

As we have the relations $m_{\rho} \simeq m_{\omega}=m$ and $\Gamma_{\rho}
\neq \Gamma_{\omega}$, the total effect is integrated over s,
\begin{equation}
\label{5} \int ds |F(s)|^{2}=1+\epsilon^{2} +2\epsilon
\cos{\theta'}
\frac{2\sqrt{\Gamma_{\omega}\Gamma_{\rho}}}{\Gamma_{\omega}+\Gamma_{\rho}}.
\end{equation}
The third term of the expression (\ref{5}), $2\epsilon
\cos{\theta'} (\frac{2\sqrt{
\Gamma_{\omega}\Gamma_{\rho}}}{\Gamma_{\omega}+\Gamma_{\rho}})$,
is the $\rho - \omega$ interference term. According to
ref.$^{\cite{Bramon}}$\footnote{$\theta=0$ in $J/\psi \rightarrow
(\rho \;{\rm or}\;\omega) \pi^{0} \rightarrow \pi^{+} \pi^{-}
2\pi^{0}$ and $\theta=\pi /2$ in $J/\psi \rightarrow (\rho \;{\rm
or}\;\omega) \pi^{0}\rightarrow \pi^{+} \pi^{-} \pi^{0}$ in the
paper of A.Bramon and J.Casulleras,Phys.Lett.,1986,{\bf 173B}:
97.}, the phase $\theta'$ is equal to zero, and the interference
effect produces a magnificent factor as a whole(we assume that the
$\omega$ resonance contribution is one),ie.
\begin{equation}
\label{6} \int ds |F(s)|^{2}=1+\epsilon^{2} +\epsilon,
\end{equation}
where $\Gamma_{\rho} \simeq 16 \Gamma_{\omega}$$^{\cite{PDG}}$ has
been used. Then, the interference between $\rho$ and $\omega$ in
the $J/\psi\rightarrow 4\pi$ provides a relation as follows
\begin{equation}\label{rel}
BR(J/\psi \rightarrow \omega (\rho) \pi^{0} \rightarrow
4\pi)=(1+\epsilon^{2}+\epsilon)BR{(J/\psi \rightarrow \omega
\pi^{0} \rightarrow 4\pi)}.
\end{equation}
The $BR(J/\psi \rightarrow \omega (\rho) \pi^{0} \rightarrow
4\pi)$ can be detected directly in experiments, but there is no a
direct experiment way to measure $BR(J/\psi \rightarrow \omega
\pi^{0} \rightarrow 4\pi)$ in the right hand side of the above
relation. Fortunately, it can be got by fitting the branch ratios
of $J/\psi \rightarrow PP$ and $PV$( where $P$ and $V$ denote
pseudoscalar- and vector mesons respectively)$^{ \cite{Bramon}}$.
As both $BR(J/\psi \rightarrow \omega (\rho) \pi^{0} \rightarrow
4\pi)$ and $BR(J/\psi \rightarrow \omega \pi^{0} \rightarrow
4\pi)$ are known, we will have $\epsilon$ by Eq.~\ref{rel} and
then obtain desired quantity $BR(\rho^{0} \rightarrow  3 \pi)$ via
Eq.~\ref{e4}.

Decays of $J/\psi$ into $(PV)$ and into $(PP)$ can be factorized
by a very simple and general consideration described as
follows$^{\cite{Bramon}}$. The decays proceed through a strongly
interacting three-gluon $(ggg)$ intermediate state and through
electromagnetic interaction mediated by one photon $\gamma$ and
$(\gamma gg)$ states. In the gluonic case, there are two $I=0$
transitions $c \bar{c} \rightarrow ggg \rightarrow \frac{(u
\bar{u}+d \bar{d})}{\sqrt{2}}$ and $c \bar{c} \rightarrow ggg
\rightarrow s \bar{s}$, which are proportional to the amplitude
$A$ and $\frac{\lambda A}{\sqrt{2}}$ respectively. The parameter
$\lambda$
 accounts for flavor-$SU(3)$
breaking effect. The electromagnetic transitions generate the
$I=1$ state $\frac{(u \bar{u}-d \bar{d})}{\sqrt{2}}$ and two $I=0$
states ($\frac{(u \bar{u}+d \bar{d})}{\sqrt{2}}$ and $s \bar{s}$).
Their amplitude are proportional to $3a,\;a$ and $-\sqrt{2}
\lambda a$ respectively. The flavor space wave functions for $P$
including $\eta$ and $\eta'$ read
\begin{equation}\label{2.6}
P=\lambda^a \Phi^a=\sqrt{2} \left(\begin{array}{ccc}
       \frac{\pi^0}{\sqrt{2}}+\frac{\eta_8}{\sqrt{6}}+{\eta_1\over
       \sqrt{3}}
            &\pi^+ &K^+   \\
    \pi^-&-\frac{\pi^0}{\sqrt{2}}+\frac{\eta_8}{\sqrt{6}}+{\eta_1\over
       \sqrt{3}}
            &K^0   \\
       K^-&\bar{K}^0&-\frac{2}{\sqrt{6}}\eta_8+{\eta_1\over
       \sqrt{3}}
       \end{array} \right),
\end{equation}
where $\eta_8=\eta \cos\theta +\eta'\sin\theta$ and $\eta_1=\eta'
\cos\theta -\eta\sin\theta$ with $\theta$ as $(\eta-\eta')$-mixing
angle, and the $V$-wave functions including $\phi$'s read
\begin{equation}\label{2.11}
   V=\sqrt{2}
\left(\begin{array}{ccc}
       \frac{\rho^0}{\sqrt{2}}+\frac{\omega}{\sqrt{2}}
            &\rho^+ &K^{*+}   \\
    \rho^-&-\frac{\rho^0}{\sqrt{2}}+\frac{\omega}{\sqrt{2}}
            &K^{*0}   \\
       K^{*-}&\bar{K}^{*0}&\phi
       \end{array} \right).
\end{equation}
With the above, the decay amplitudes of $(J/\psi \rightarrow
PP,\;PV)$ are as follows
\begin{equation}
\label{7a} {A(\pi^{+} \pi{-})}=3a,
\end{equation}
\begin{equation}
\label{7b} A(K^{+} K^{-})=\frac{1}{2}(1-\lambda)A+(2+\lambda)a,
\end{equation}
\begin{equation}
\label{7c} A(K^{0}
\bar{K^{0}})=\frac{1}{2}(1-\lambda)A-(1-\lambda)a,
\end{equation}
\begin{equation}
\label{7d} A(\rho^{0} \pi^{0})=f_{v}(A+a),
\end{equation}
\begin{equation}
\label{7e} A(K^{\star{+}}
K^{-})=f_{v}[\frac{1}{2}(1+\lambda)A+(2-\lambda)a],
\end{equation}
\begin{equation}
\label{7f} A(K^{\star{0}} \bar{K^{0}})=f_{v}
[\frac{1}{2}(1+\lambda)A-(1+\lambda)a],
\end{equation}
\begin{equation}
\label{7g} A(\omega \pi^{0})=f_{v}(3a),
\end{equation}
\begin{equation}
\label{7h} A(\rho \eta^{\prime})=f_{v}(3a X_{\eta^{\prime}}),
\end{equation}
\begin{equation}
\label{7i}
A(\omega\eta^{\prime})=f_{v}[(A+a)X_{\eta^{\prime}}+\sqrt{2}
rA(\sqrt{2}X_{\eta^{\prime}}+Y_{\eta^{\prime}}],
\end{equation}
\begin{equation}
\label{7j} A(\rho \eta)=f_{v}(3a X_{\eta}),
\end{equation}
\begin{equation}
\label{7k} A(\omega\eta)=f_{v}[(A+a)X_{\eta}+\sqrt{2}
rA(\sqrt(2)X_{\eta}+Y_{\eta}],
\end{equation}
\begin{equation}
\label{7l} A(\phi \eta)=f_{v}[(A-2a) \lambda Y_{\eta}+rA(\sqrt{2}
X_{\eta}+Y_{\eta}) \frac{(1+\lambda)}{2}],
\end{equation}
\begin{equation}
\label{7m} A(\phi \eta^{\prime})=f_{v}[(A-2a) \lambda
Y_{\eta^{\prime}}+ rA(\sqrt{2}
X_{\eta^{\prime}}+Y_{\eta^{\prime}}) \frac{(1+\lambda)}{2}],
\end{equation}
where $X_{\eta}=\sqrt{\frac{1}{3}}\cos \theta
-\sqrt{\frac{2}{3}}\sin \theta$,
$X_{\eta^{\prime}}=\sqrt{\frac{1}{3}}\sin\theta+\sqrt{\frac{2}{3}}\cos
\theta$,
$Y_{\eta}=-X_{\eta^{\prime}}$,$Y_{\eta^{\prime}}=X_{\eta}$. The
additional parameter $r$ is the relative weight of the
disconnected diagram to connected diagram for the decays involving
the final state $\eta$ or
$\eta^{\prime}$$^{\cite{Jousset1}\cite{Jousset2}}$. The
Eqs.~\ref{7a}-~\ref{7g} are same as ones in
ref.$^{\cite{Bramon}}$, and others are new.

The corresponding branching ratios of these decays are following
\begin{equation}
\label{8a} Br(\pi^{+} \pi^{-})=9a^{2},
\end{equation}
\begin{equation}
\label{8b} Br(K^{+}K^{-})=|\frac
{1}{2}(1-\lambda)A+(2+\lambda)ae^{i \phi}|^{2},
\end{equation}
\begin{equation}
\label{8c} Br(K^{0} \bar{K^{0}}) =|\frac {1}{2}(1-\lambda)A-
(1-\lambda)ae^{i \phi}|^{2},
\end{equation}
\begin{equation}
\label{8d} Br(\rho^{0} \pi^{0})=f^{2}_{v}|(A+ae^{i \phi})|^{2},
\end{equation}
\begin{equation}
\label{8e} Br(K^{\star{+}} K^{-})=f^{2}_{v}|\frac{1}{2}(1+
\lambda)A+(2-\lambda)ae^{i \phi}|^{2},
\end{equation}
\begin{equation}
\label{8f} Br(K^{\star{0}} \bar{K^{0}})=f^{2}_{v}|\frac {1}{2}(1+
\lambda)A-(1+ \lambda)ae^{i \phi}|^2,
\end{equation}
\begin{equation}
\label{8g}
Br(\omega(\rho^0)\pi^0)=(1+\epsilon+\epsilon^{2})f^{2}_{v}9a^{2},
\end{equation}
\begin{equation}
\label{8h} Br(\rho \eta^{\prime})=f^{2}_{v}|3a
X_{\eta^{\prime}}|^{2},
\end{equation}
\begin{equation}
\label{8i}
Br(\omega\eta^{\prime})=f^{2}_{v}|(A+ae^{i\phi})X_{\eta^{\prime}}+\sqrt{2}
rA(\sqrt{2}X_{\eta^{\prime}}+Y_{\eta^{\prime}}|^{2},
\end{equation}
\begin{equation}
\label{8j} Br(\rho \eta)=f^{2}_{v}|3a X_{\eta}|^{2},
\end{equation}
\begin{equation}
\label{8k}
Br(\omega\eta)=f^{2}_{v}|(A+ae^{i\phi})X_{\eta}+\sqrt{2}
rA(\sqrt(2)X_{\eta}+Y_{\eta}|^{2},
\end{equation}
\begin{equation}
\label{8l} Br(\phi \eta)=f^{2}_{v}|(A-2ae^{i\phi}) \lambda
Y_{\eta}+rA(\sqrt{2} X_{\eta}+Y_{\eta}) \frac{1+\lambda}{2}|^{2},
\end{equation}
\begin{equation}
\label{8m} Br(\phi \eta^{\prime})=f^{2}_{v}|(A-2ae^{i\phi})
\lambda Y_{\eta^{\prime}}+ rA(\sqrt{2}
X_{\eta^{\prime}}+Y_{\eta^{\prime}}) \frac{1+\lambda}{2}|^{2},
\end{equation}
where $\phi$ is their relative phase between $A$ and $a$, and the
parameter $A$ and $a$ are  real. In the Eq.~\ref{8d} the $\rho -
\omega$ interference effect is subtracted from the branching ratio
of $J/\psi \rightarrow \rho \pi^{0}$. In the Eq.~\ref{8g}, a
 magnificent factor $1+\epsilon+\epsilon^{2}$ has been added due to
Eq.~\ref{rel} in order to taking  the $\rho - \omega$ interference
effects into account. Actually, through directly detecting the
data of $J/\psi\rightarrow 4\pi$ one can only get
$Br(\omega(\rho^0)\pi^0)$ rather than $Br(\omega \pi^0)$ which is
equal to $f^{2}_{v}9a^{2}$.

In the  branching ratio formulae of Eqs.~\ref{8a}-~\ref{8m} we do
not write out the corresponding phase-space factors explicitly
which are proportional to the cube of the final momenta in
two-body decays. They will be taken into account in the practical
phenomenological fit later.

\section{Datum fit to obtain
$BR(\rho \rightarrow \pi^{+}\pi^{-}\pi^{0})$}

\normalsize
~~~~Let's now use the data in PDG-2002$^{\cite{PDG}}$
to perform the fit to all branching ratios of
Eq.~\ref{8a}-~\ref{8m}. This fit will lead to determining the
$\epsilon$ and $BR(\rho \rightarrow \pi^{+}\pi^{-}\pi^{0})$. The
experimental branching ratio data of PDG-2002 are listed in the
second column of Table $1$.

Firstly, following ref.$^{\cite{Bramon}}$, we use the branch ratio
data of ($J/\psi \rightarrow PP,\;V(\pi^0,K)$) only to perform a
fit to Eqs.~\ref{8a}-~\ref{8g}(call it as $(PP,V(\pi^0,K))-$fit
hereafter). In this case, there are seven equations with six
adjustable free parameters
$a,\;A,\;\lambda,\;\phi,\;f_V,\;\epsilon$, and hence it is an
over-determination problem with potential of predictions. The fit
with minimum $\chi^{2}=0.46$ leads to the values of the parameters
and seven corresponding branching ratios listed in the third
column of Table 1, in which $(\omega \pi^{0})_{uncor}$ and
$(\omega \pi^{0})_{cor}$ represent $Br(J/\psi \rightarrow \omega
(\rho)\pi^0 \rightarrow 4\pi)$ and $Br(J/\psi \rightarrow \omega
\pi^0 \rightarrow 4\pi)$ respectively, i.e.,
\begin{equation}\label{cor}
Br(J/\psi \rightarrow \omega \pi^{0})_{cor}=f^{2}_{v}9a^{2}.
\end{equation}
 In the fit (see Table 1), we have $(\omega
\pi^{0})_{uncor}=(4.2 \pm 0.61 )\times 10^{-4}$, $a=0.21\pm 0.02$,
$A=2.94\pm 0.72$ and the interference factor $\epsilon=0.71\pm
 0.58$, then we obtain $(\omega \pi^{0})_{cor}=f^{2}_{v}9a^{2}=(1.89 \pm 0.83)
\times 10^{-4}$. In other hand, from
Eqs.(\ref{e4})(\ref{7d})(\ref{7g}), $\epsilon$ reads
\begin{equation}
\epsilon=\frac{|A+ae^{i \phi}|}{|3a|} \times \sqrt{\frac{Br(\rho
\rightarrow 3\pi)}{Br(\omega \rightarrow  3\pi)}}.
\end{equation}
Then the branching ratio of $\rho \rightarrow \pi^{+} \pi^{-}
\pi^{0}$ is predicted as follows
\begin{equation}
\label{calr3p}
Br(\rho \rightarrow 3\pi)=Br(\omega \rightarrow 3\pi)\times (\frac{3|a|}
{|A+ae^{i \phi}|})^{2} \epsilon^{2}.
\end{equation}
Substituting the $a-,\;A-$ and $\phi-$values obtained from the fit
(see the third column of Table 1) and experiment data of
$Br(\omega \rightarrow 3\pi)$ into  Eq.~\ref{cor} and
Eq.~\ref{calr3p}, we then obtain the $(PP,V(\pi^0,K))-$fit's
results as follows
\begin{eqnarray}\label{result1a}
Br(J/\psi\rightarrow \omega
\pi^{0})_{cor}|_{(PP,V(\pi^0,K))}&=&(1.89 \pm 0.83) \times
10^{-4},\\
\label{result1b}
 Br(\rho \rightarrow \pi^{+} \pi^{-}
\pi^{0})|_{(PP,V(\pi^0,K))}&=&(2.0 \pm 1.64) \times 10^{-2}.
\end{eqnarray}
 \vskip 3mm
\begin{center}
\small
\noindent TABLE 1: The second column displayed the
experimental values for the branch ratios\\
of $J/\psi \rightarrow PP$ and $J/\psi \rightarrow PV$
in PDG-2000 datum . The results of a fit to the\\
first seven branching ratios is listed in the third
column.The results of a fit to the\\
total thirteen branching ratios is listed in the forth column.
\end{center}
\begin{center}
\small
\begin{tabular}{|l|l|l|l|l|}\hline \hline
$J/\psi decay$       &PDG-2002$(10^{-4})$ & a partial fit1$(10^{-4})$   & a global fit2$(10^{-4})$ \\\hline
$1.\pi^{+} \pi^{-}$           & $1.47 \pm 0.23$  & $1.44  \pm 0.23$
 & $1.92  \pm 0.05$  \\
$2.K^{+} K^{-}$               & $2.37 \pm 0.31$  & $2.45  \pm 0.28$
 & $2.04  \pm 0.08$  \\
$3.K^{0} \bar{K^{0}}$         & $1.08  \pm 0.14$ & $1.06  \pm 0.14$
 & $0.87  \pm 0.05$  \\
$4.\rho^{0} \pi^{0}$          & $42.0  \pm 5$    & $43.04 \pm 4.48$
 & $41.97 \pm 0.68$  \\
$5.K^{\star{+}} K^{-}$        & $25.0  \pm 2.0$  & $24.11 \pm 1.41$
 & $23.64  \pm 0.5$  \\
$6.K^{\star{0}} K^{0}$        & $ 21.0 \pm 2.0$  & $21.66 \pm 1.7$
 & $24.21 \pm 0.49$  \\
$7.(\omega \pi^{0})_{uncor}$  & $4.2   \pm 0.6$  & $4.2   \pm 0.61$
 & $4.2  \pm 0.2$  \\
$8.(\rho \eta^{\prime})$      & $1.05  \pm 0.18$ &
 & $0.7  \pm 0.05$  \\
$9.(\omega \eta^{\prime})$    & $1.67  \pm 0.25$ &
 & $1.73 \pm 0.11$  \\
$10.(\rho \eta)$              & $1.93  \pm 0.23$ &
 & $1.82 \pm 0.08$  \\
$11.(\omega \eta)$            & $15.8  \pm 1.6$  &
 & $18.32 \pm 0.36$  \\
$12.(\phi \eta)$              & $6.5   \pm 0.7$  &
 & $5.85 \pm 0.23$   \\
$13.(\phi \eta^{\prime})$     & $3.3   \pm 0.4$  &
 & $2.55 \pm 0.23$   \\\hline
$\chi^{2}$  &    & $0.46/1$           &   $21.4/5$            \\
  EDM       &    & $0.45E-06$         &   $0.69E-06$          \\\hline
   fit $a$  &    & $0.21 \pm 0.02$    &   $0.24 \pm 0.012$    \\
       $A$  &    & $2.94 \pm 0.72$    &   $2.69 \pm 0.17$    \\
  $\lambda$ &    & $0.6  \pm 0.1$     &   $0.62 \pm 0.03$    \\
   $\phi$   &    & $1.37  \pm 0.14$   &   $1.6  \pm 0.11$    \\
   $f_{v}$  &    & $1.26 \pm 0.36$    &   $1.38 \pm 0.1 $    \\
 $\epsilon$ &    & $0.71 \pm 0.58$    &   $0.3  \pm 0.16$    \\
 $\theta$   &    &                  &   $-0.343\pm0.026 $   \\
 $r$        &    &                  &   $-0.144\pm 0.001$   \\\hline
$(\omega \pi^{0})_{cor}$ &  & $(1.89 \pm 0.83)\times 10^{-4}$ &
$(3.02\pm 0.2)\times 10^{-4}$
\\\hline $\rho \rightarrow 3\pi$ &  & $(2.0 \pm 1.64)\times
10^{-2}$ & $(0.59\pm 0.315)\times 10^{-2}$
\\\hline \hline
\end{tabular}
\end{center}
\vskip 3mm

Secondly, we perform more complete datum fit in which the
processes of $J/\psi \rightarrow V \eta$ and $J/\psi \rightarrow V
\eta'$ are included. In this case, there are 13 equations
(\ref{8a}-\ref{8m}) and eight free parameters:
$a,\;A,\;\lambda,\;\phi,\;f_V,\;\epsilon,\;\theta,\;r$. And hence
it is an over-determination problem with more constraints, and
will be called as $(PP,V(\pi^0,K,\eta,\eta'))-$fit hereafter. The
results are as follows
\begin{eqnarray}\label{result2a}
Br(J/\psi\rightarrow \omega
\pi^{0})_{cor}|_{(PP,V(\pi^0,K,\eta,\eta'))}&=&(3.02\pm 0.2)
\times
10^{-4},\\
\label{result2b}
Br(\rho \rightarrow \pi^{+} \pi^{-}
\pi^{0})|_{(PP,V(\pi^0,K,\eta,\eta'))}&=&(0.59\pm  0.315)\times 10^{-2},\\
\label{result2c} \theta=-0.343\pm0.026&=&-19.68^o\pm 1.49^o,
\end{eqnarray}
where $\eta-\eta'-$mixing angle $\theta$ is agreement with one in
ref.$^{\cite{Jousset1}\cite{Jousset2}}$, and both
$Br(J/\psi\rightarrow \omega \pi^{0})_{cor}$ and $Br(\rho
\rightarrow \pi^{+} \pi^{-} \pi^{0}) $ are reasonable agreement
with the results (\ref{result1a})(\ref{result1b}) obtained by
$(PP,V(\pi^0,K))-$fit within the errors.

 The parameter $\lambda$ is the constituent quark
mass ratio $m_u/m_s$ which should be about
0.6$^{\cite{Jousset2}\cite{balt1}\cite{balt2}}$ due to light
flavor SU(3)-breaking. The results of $\lambda \simeq 0.6\pm 0.1$
for $(PP,V(\pi^0,K))-$fit and $\lambda \simeq 0.62\pm 0.03$ for
$(PP,V(\pi^0,K,\eta,\eta'))-$fit indicate the fits meet this
requirement, and, hence, the results yielded by them are rather
reliable.

\section{Large isospin breaking effect in decay
$\rho^{0} \rightarrow \pi^{+} \pi^{-} \pi^{0}$}
 \vskip 3mm
~~~~In this section, following ref.$^{\cite{WJY}}$, we provide a
theoretical estimation to $Br(\rho^{0} \rightarrow \pi^{+} \pi^{-}
\pi^{0})$. Using Feynman propagators method, the on-shell
amplitude$^{\cite{WJY}}$ of the decay $\rho \rightarrow \pi^{+}
\pi^{-} \pi^{0}$ is determined by
\begin{equation}\label{1}
{\cal M}_{\rho^{0}\rightarrow3\pi}=
\Big(f_{\rho3\pi}+\frac{\Pi_{\rho\omega}(p^{2})f_{\omega3\pi}}{p^{2}-
m^{2}_{\omega}+im_{\omega}\Gamma_{\omega}}\Big)\Big|_{p^{2}=m^{2}_{\rho}},
\end{equation}
where, the momentum-dependent $\rho^{0} - \omega$ interference
amplitude $\Pi_{\rho \omega}(q^{2})$ is defined by the
$\rho-\omega$ interaction Lagrangian ${\cal L}_{\rho\omega}$ as
follows
\begin{equation}\label{2}
{\cal L}_{\rho \omega}=\int {\frac{d^{4}p}{(2 \pi)^{4}} e^{-ip
\cdot x } \Pi_{ \rho \omega}(p^{2}) (g^{\mu
\nu}-\frac{p^{\mu}p^{\nu}} {p^{2}})
\omega_{\mu}(p)\rho^0_{\nu}(x)}.
\end{equation}
The first and second term of expression (\ref{1}) correspond to
the contributions of direct coupling $(\rho^0-3\pi)$ and $\omega
\-$ resonance exchange respectively. Because $m_{\rho} \simeq
m_{\omega}$ and $\Gamma_{\omega}$ is small,
 the denominator of the second term is small. Therefore, contribution
from $\omega$ resonance exchange is large. This is called "hidden
isospin symmetry breaking effect" according to $^{\cite{WJY}}$.
This effect brings a significant contribution and plays an
essential role in the decay $\rho \rightarrow \pi^{+} \pi^{-}
\pi^{0}$. So when we deal with the decay of $\rho \rightarrow
\pi^{+} \pi^{-} \pi^{0}$, the process $\rho \rightarrow \omega
\rightarrow \pi^{+} \pi^{-} \pi^{0}$ must be considered.

In fact, the contributions of $\rho^{0} - \omega$ interference are
dominant and the direct coupling can be omitted. The direct
coupling $f_{\rho3\pi}\propto (m_d-m_u)$, therefore, it is very
small. In order to be sure of this point, we derive this quantity
in a practical model called as $U(2)_L\times U(2)_R$ chiral theory
of mesons $^{\cite{Li95}}$ in follows. Denoting the direct
vertices of $\rho^{0} - 3\pi$ as $ {\cal L}_{\rho 3\pi}=f_{\rho
3\pi}\epsilon^{\mu\nu\alpha\beta}\epsilon_{ijk} \rho^0_\mu
\partial_\nu \pi^i \partial_\alpha \pi^j \partial_\beta \pi^k$,
 then $f_{\rho3\pi}$ can be calculated in this theory$^{\cite{Li95}}$ and has
the form
\begin{eqnarray}\label{3.3}
f_{\rho 3\pi}=-\frac{m_d-m_u}{\pi^2gf_\pi^3m}
\Big(1-\frac{16c}{3g}+\frac{6c^2}{g^2}-\frac{8c^3}{3g^3}\Big)\sim-2\times
10^{-11}{\rm MeV}^{-3},
\end{eqnarray}
where the values of model's parameters $m,\;g,\;c$ determined in
ref.$^{\cite{Li95}}$ have been used. To the second term in the
parentheses of expression (\ref{1}),
$\Pi_{\rho\omega}(m_\rho^{2})$ has be determined to approximate
$-4\times 10^3{\rm MeV}^2$ $^{\cite{CB,GaoYanRho}}$.
$\omega\rightarrow 3\pi$ is the dominant channel for
$\omega-$decays, and hence $f_{\omega3\pi}$ can be estimated by
using the width $\Gamma_{\omega\rightarrow3\pi}=7.5$MeV. It's
approximate value is about $3\times 10^{-7}{\rm MeV}^{-3}$. Thus
the typical value of the second term in expression (\ref{1}) is
$(5+2i)\times 10^{-8}{\rm MeV}^{-3}$ approximately. Comparing it
with expression (\ref{3.3}), we can see that the direct coupling
$f_{\rho 3\pi}$ is indeed very small, and it is ignorable.
Therefore, discarding $f_{\rho 3\pi}$ in expression (\ref{1}), we
have, approximately,
\begin{equation}\label{4}
\Gamma_{\rho^{0}\rightarrow3\pi}=
\Big|\frac{\Pi_{\rho\omega}(m_\rho^{2})}{m_\rho^{2}-
m^{2}_{\omega}+im_{\omega}\Gamma_{\omega}}\Big|^2\Gamma_{\omega\rightarrow3\pi}.
\end{equation}
This equation means that the contributions due to
$\rho-\omega-$interference to $Br(\rho^0 \rightarrow 3\pi)$ are
dominate, or the hidden isospin-breaking effects introduced in
$^{\cite{WJY}}$ are dominate for the process $\rho^{0}\rightarrow
3\pi.$ From expression (\ref{4}) we obtain desired result as
follows
\begin{equation}\label{f}
BR(\rho^0\rightarrow 3\pi)\simeq 0.2\times 10^{-2}.
\end{equation}
Our experiment datum fitting result (\ref{result2b}) is consistent
with this theoretical estimation result. This fact indicates that
 both $(PP,V(\pi^0,K))-$fit and $(PP,V(\pi^0,K,\eta,\eta'))-$fit
are reasonable even though the resulting $BR(\rho^0\rightarrow
3\pi)$ is much larger than one in ref.$^{\cite{Vas}}$ and rather
closes the upper limit for it in ref.$^{\cite{Abra}}$.

\section{Discussion} \vskip 2mm \normalsize

~~~~Through the study presented in the above, we conclude that
$\rho-\omega-$interference effects can be detected in the $J/\psi
\rightarrow \pi^+\pi^-\pi^0\pi^0$ decay, which receives a
contribution from the $\rho^0\rightarrow \pi^+\pi^-\pi^0$ decay
mode. $J/\psi$ decays offer an almost unique opportunity for
observing $\rho^0\rightarrow \pi^+\pi^-\pi^0$, where the smallness
of $Br(\rho^0\rightarrow \pi^+\pi^-\pi^0)/Br(\omega\rightarrow
\pi^+\pi^-\pi^0)$ is compensated by the large ratio $A(J/\psi
\rightarrow \rho\pi^0)/A(J/\psi \rightarrow \omega\pi^0)$ between
a (simply Zweig-forbidden) strong amplitude over an EM one. This
is the key point for the practical determining $Br(\rho^0
\rightarrow \pi^+\pi^-\pi^0)$ through employing $J/\psi$ decay
branching radios. Our results for 2 datum-fits are $Br(\rho
\rightarrow \pi^{+} \pi^{-} \pi^{0})|_{(PP,V(\pi^0,K))}=(2.0\pm
1.64)\times 10^{-2}$ and $Br(\rho \rightarrow \pi^{+} \pi^{-}
\pi^{0})|_{(PP,V(\pi^0,K,\eta,\eta'))}=(0.59\pm  0.315)\times
10^{-2}$ respectively, which are anomalously large and match each
other within the errors.

In order to pursue whether these anomalously large results of
$Br(\rho \rightarrow \pi^{+} \pi^{-} \pi^{0})$ are reasonable or
not, a theoretical estimation for $\rho-\omega-$interference
effects to the process of $(\rho \rightarrow \pi^{+} \pi^{-}
\pi^{0})$ has also been discussed in this paper. Following
ref.$^{\cite{WJY}}$, we found that the contributions due to so
called hidden isospin-breaking effects are dominate for the
process $\rho \rightarrow \pi^{+} \pi^{-} \pi^{0}$. The
theoretical prediction is $BR(\rho^0\rightarrow 3\pi)\simeq
0.2\times 10^{-2}$ which is in good agreement with our datum-fit
results. Then, considering this fact and noting that both result
of $\eta-\eta'-$angle $\theta$ and the result of constituent quark
ratio $\lambda=m_u/m_s$ obtained by the fits are also reasonable,
we conclude that $Br(\rho \rightarrow \pi^{+} \pi^{-} \pi^{0})\sim
10^{-3}-10^{-2}$ is reliable.

Finally, we like to argue that in order to reduce the error-bar of
$Br(\rho \rightarrow \pi^{+} \pi^{-} \pi^{0})$, more precisely
experimental measurements to $(J/\psi \rightarrow PP,\;PV)$ are
expected. The high quality data for $J/\psi$ in the future BESIII
would be useful.

\vskip 5mm \indent We would like to thank Zheng Zhi-Peng, Shen
Xiao-Yan, Zhu Yucan, Yuan Chang-zheng,  Fang Shuang-shi for
helpful discussions. This work is partially supported by NSF of
China 90103002 and the Grant of National Laboratory at the
Institute of High Energy Physics, Beijing.

\end{document}